\documentclass[a4paper,fleqn]{cas-sc}
\usepackage{placeins}
\usepackage{lineno}
\usepackage{setspace}
\usepackage{algorithm}
\usepackage{algpseudocode}
\usepackage{graphicx}
\usepackage{subcaption}
\usepackage{xcolor} 
\usepackage{tikz}
\usetikzlibrary{arrows.meta, positioning}
\usepackage{pgfplots}
\usepackage{float}
\usepackage{geometry}
\usepackage{amsmath}
\usepgfplotslibrary{colormaps}
\usepackage{tcolorbox}
\usepackage[authoryear]{natbib}
\usepackage{bbm}
\usepackage{enumitem}
\def\tsc#1{\csdef{#1}{\textsc{\lowercase{#1}}\xspace}}
\tsc{WGM}
\tsc{QE}
\newcommand{\cyr}[1]{\textcolor{black}{#1}}

\begin{document}
\let\WriteBookmarks\relax
\def\floatpagepagefraction{1}
\def\textpagefraction{.001}

\shorttitle{$\gamma$-Index \& Surface Solar Irradiance}    
\title[mode=title]{A Tolerance-Based Framework for Spatio-Temporal Forecast Validation Using the $\gamma$-Index}

\shortauthors{C. Voyant}  
	\author[label1]{Cyril Voyant} [orcid=0000-0003-3144-5945]
	 \cormark[1]
	 \ead{cyril.voyant@minesparis.psl.eu}
\affiliation [label1] {organization={OIE Laboratory},
            addressline={Mines-PSL}, 
           city={Sophia-Antipolis},
            postcode={F-06904}, 
            state={Antibes},
            country={France}}
\cortext[cor1]{Corresponding author}

\begin{abstract}
\cyr{Classical field forecast evaluation relies mainly on local scores such as \texttt{RMSE} or \texttt{MAE}. These metrics severely over penalize small spatial or temporal displacements of coherent structures, a limitation known as the double penalty issue and common to many forecasting domains. The present paper introduces a tolerance based framework built on the three dimensional $\gamma$-Index, initially designed for medical dose verification, as a unified acceptance criterion for gridded forecasts. The method embeds explicit margins in space (\texttt{DTA}), time (\texttt{TTA}) and intensity (\texttt{IDT}), and evaluates whether predictions agree with observations within predefined physical bounds rather than through pixel wise differences only. A synthetic illustration is first used to show why conventional metrics can misrepresent usable forecasts. The approach is then applied to satellite derived \texttt{SSI} fields to demonstrate operational behaviour on a real dataset. Results confirm that the $\gamma$ criterion preserves structural consistency under minor positional noise while isolating physically significant discrepancies. The formulation is generic and can be implemented for any gridded variable provided meaningful tolerances are defined, offering a pragmatic complement to existing spatial verification tools in general forecasting workflows.}
\end{abstract}

\begin{graphicalabstract}
        \centering
        \includegraphics[width=1\textwidth]{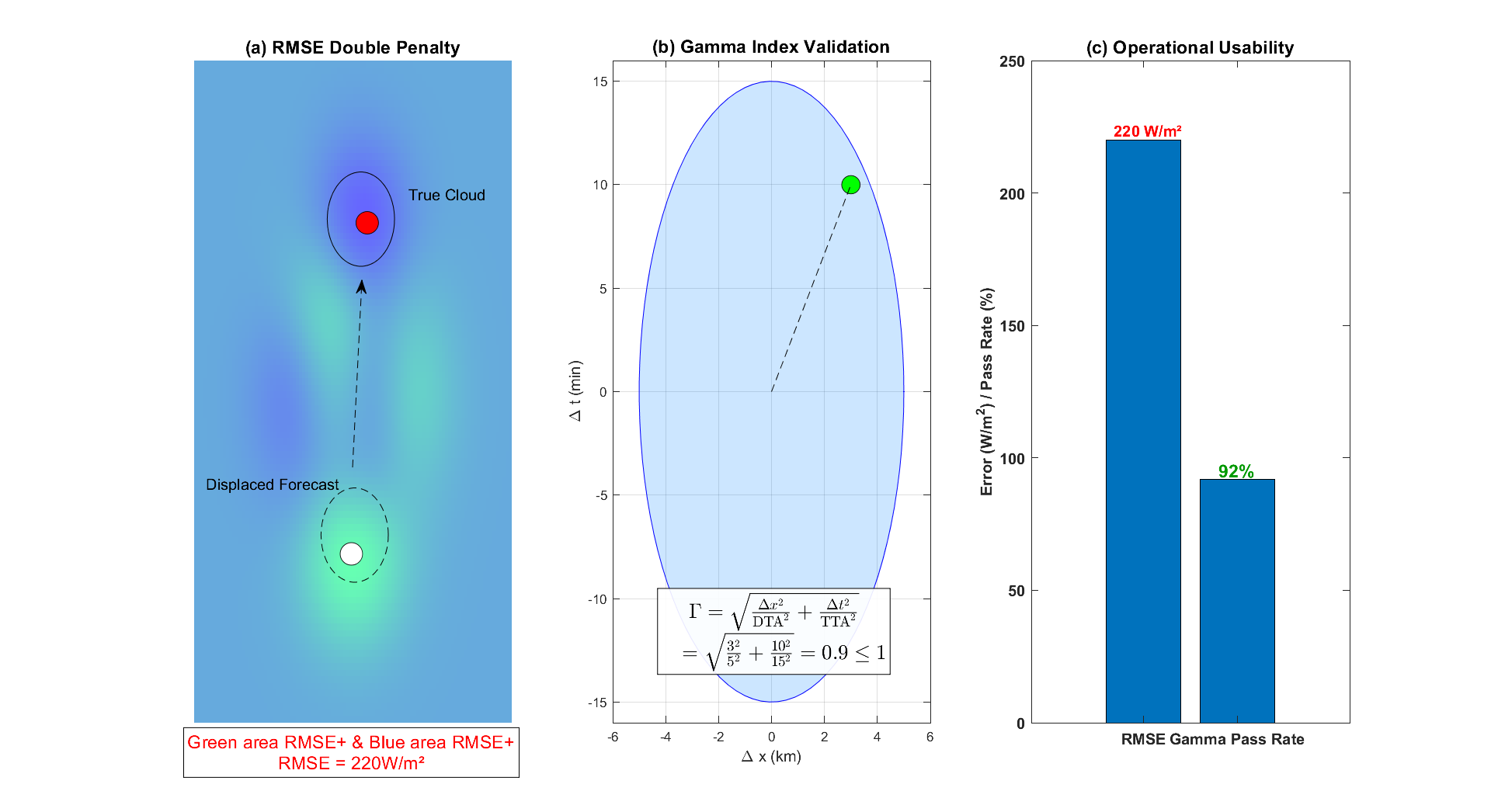}
        \label{fig:gamma_abstract}
\begin{tcolorbox}[colframe=black, colback=gray!10, sharp corners=south]
    \textbf{(a)~\texttt{RMSE} critical flaw}: A cloud displaced by 5\,km (spatial) and 10\,min (temporal) incurs dual penalties, yielding misleadingly high error (220\,W/m\textsuperscript{2}).  
    
    \textbf{(b)~$\gamma$-Index solution}: Validation via  $\Gamma = \sqrt{\frac{\Delta x^2}{\texttt{DTA}^2} + \frac{\Delta t^2}{\texttt{TTA}^2}} \leq 1$  
    with tolerances (\texttt{DTA} = 5\,km, \texttt{TTA} = 15\,min). Displacements within thresholds are deemed acceptable.  
    
    \textbf{(c)~Operational impact}: 92\% of displaced forecasts are validated as usable (Gamma) versus high \texttt{RMSE} value inducing significant failure rate.  
\end{tcolorbox}
\end{graphicalabstract}

\begin{highlights}
\item First application of the 3D $\gamma$-Index to spatio–temporal forecast validation, addressing the double penalty issue inherent to traditional metrics such as \texttt{RMSE}. 
\item Assessment based on a decade of high-resolution satellite derived datasets ($\sim$2\,km) over Corsica (\texttt{HelioClim-3}, \texttt{Heliostat}).  
\item Shift from pixel wise error metrics to a tolerance based validation framework applicable to any gridded forecast field.  
\item Definition of $\gamma$-based performance thresholds supporting interpretable, operationally consistent model benchmarking.  
\end{highlights}


\maketitle
%
%
\section{Introduction}
\cyr{Comparing maps represents more than a graphical exercise. A forecast field describes spatial organisation of physical processes, so validation must assess location, motion, and amplitude jointly \citep{Casati2004,bbbb}. In meteorology, many variables, including clouds, precipitation or wind, exhibit strong spatial coherence and sharp gradients. Small errors in translation may produce large pixel wise penalties and mask realistic dynamics \citep{ebert2000,Roberts2008,Gilleland2009}. Gridded irradiance forecasts drive regional production estimates and dispatch planning in energy systems. Pixel wise errors may be misinterpreted as severe drops, leading to conservative operational actions and unnecessary corrective measures when the field dynamics are realistic but slightly shifted \citep{YANG202020,VOYANT1}. Therefore, classical scores such as \texttt{RMSE} and \texttt{MAE} may mask structural skill and distort verification for displaced coherent patterns \citep{YANG201860}. Map based verification requires physics aware criteria capable of distinguishing displacement uncertainty from genuine discrepancies, a need also emphasized in spatial verification research \citep{davis2006,morales2015,aaaa,cccc}.}

\subsection{Forecast Verification for Gridded Fields}
\cyr{Evaluation of gridded forecasts still relies mainly on local scores such as \texttt{RMSE}, \texttt{MAE}, \texttt{MBE}, $R^2$, or Willmott's $d$-index \citep{Willmott1981}. These metrics assume exact spatial and temporal coincidence between prediction and reference data. This assumption may be violated for many field forecasting situations, so verification based on point wise differences can mask realistic agreement. To reduce this limitation, the literature proposes neighbourhood or fuzzy approaches that compare local summaries over spatial windows \citep{Roberts2008}. Object based methods extract coherent features and compare their attributes, which supports a more physical interpretation of misses, false alarms and structure errors \citep{ebert2000,davis2006,morales2015,aaaa,bbbb}. Scale separation techniques evaluate skill across spatial resolutions and are now standard for precipitation type fields, but they primarily describe performance by scale rather than providing a single operational acceptance criterion \citep{Casati2004,Gilleland2009}. In parallel, structural similarity metrics developed in image analysis, such as \texttt{SSIM}, \texttt{FSIM}, \texttt{GMSD} and \texttt{VIF}, aim at preserving perceptual or geometric fidelity \citep{Mason2020}. While useful for image quality assessment, their objective is not aligned with forecast validation when users require explicit physical margins in space, time and intensity \citep{Peng2020,Safari2023,Aggarwal2019}. Overall, the above approaches remain more demanding to interpret for a general forecasting audience and they do not provide an explicit unified acceptance criterion in physical units.}

\subsection{Gap, Problem Statement, and Proposed Contribution}
\cyr{Although effective, the cited spatial and structural frameworks remain oriented toward diagnostics by window or by scale, and they do not yield a unique field acceptance rule. Consequently, interpretation of \texttt{RMSE} or \texttt{MAE} for displaced coherent patterns remains ambiguous. More importantly, the surveyed methods do not provide any classical metric with an explicit acceptance criterion combining spatial displacement and temporal offset with intensity deviation in physical units under a single threshold. This absence defines the research gap and the forecasting problem addressed here. It motivates the following problem statement: how to evaluate gridded forecasts in a way that (i) remains robust to small, operationally acceptable displacements, while (ii) still detects genuinely significant discrepancies in amplitude and structure. To respond to this gap, the paper proposes the three dimensional $\gamma$-Index as a tolerance based acceptance framework, calibrated in the units of the variable and directly comparable with classical scores \citep{Low1998}. This parameter embeds explicit tolerances in space (\texttt{DTA}), time (\texttt{TTA}), and intensity (\texttt{IDT}) into a single normalized measure, and assesses whether forecast and observed fields agree within predefined physical bounds. Previous works \citep{VOYANT1,VOYANT2} have shown its potential in renewable energy forecasting. The present study extends this concept to the spatio--temporal validation of Surface Solar Irradiance (\texttt{SSI}) fields and documents its behaviour relative to classical local metrics such as \texttt{RMSE} and \texttt{MAE}.}

\cyr{Section~2 describes the $\gamma$-Index formulation and the practical parameterisation for gridded forecasts. Section~3 presents a synthetic illustration and then the application to \texttt{SSI} fields over Corsica. Section~4 discusses interpretation, practical use, and limitations, including tolerance selection. Section~5 concludes and outlines prospects for broader forecasting applications.}
%
%
\section{Materials and Methods}\label{sec:Methods}
\cyr{Map forecast verification requires an explicit and reproducible acceptance rule. The objective of this section is to describe the mathematical definition of the $\gamma$-Index and the numerical workflow used in the paper. The workflow follows the original formulation introduced in medical physics for three dimensional dose comparison \citep{Low1998}. }

\subsection{Introduction of the $\gamma$-Index}
\cyr{Classical error scores evaluate forecasts through point wise differences at fixed locations. The same assumption can produce inflated penalties when predicted and observed structures are physically consistent but shifted in space or time \citep{YANG202020,Casati2004,Roberts2008,Gilleland2009}. The $\gamma$-Index introduces predefined tolerances in spatial distance \texttt{DTA}, temporal offset \texttt{TTA}, and intensity deviation \texttt{IDT}. The metric quantifies whether forecast values agree with observations within these physical margins. The metric maps both forecast and observations into a normalized $\mathbb{R}^3$ tolerance space, where location, time, and intensity are treated as components of a single distance.}

\subsection{Metric Definition and Numerical Workflow}
\begin{figure}
        \centering
        \includegraphics[width=0.8\textwidth]{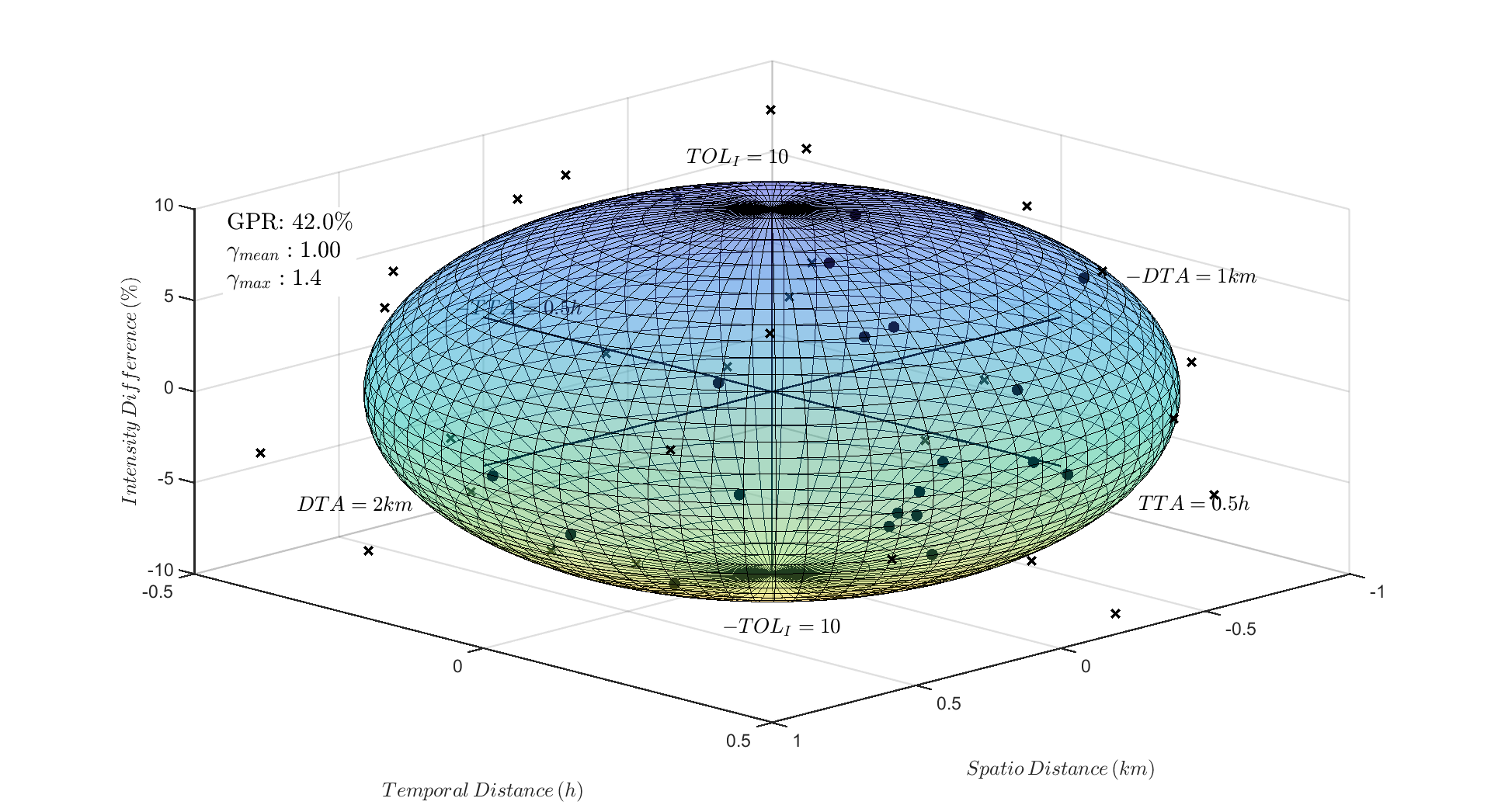}
        \caption{3D representation of the $\gamma$-Index tolerance region with \texttt{DTA}, \texttt{TTA}, and \texttt{IDT}. The black circles represent points inside the tolerance region ($\gamma \leq 1$), while black crosses represent points outside ($\gamma > 1$).}
        \label{fig:gamma_ellipse}
    \end{figure}

\cyr{The tolerance based formulation can be applied to any gridded forecast variable, provided that physically meaningful spatial (\texttt{DTA}), temporal (\texttt{TTA}), and intensity (\texttt{IDT}) tolerances are defined \citep{VOYANT1,VOYANT2,Roberts2008}. The $\gamma$-Index defines an acceptance rule based on operational thresholds. Its boundedness through the passing rate \texttt{GPR} and its non linear dependence on joint discrepancies differ from classical point wise scores such as \texttt{RMSE}, \texttt{MAE}, or \texttt{MBE}. For a prediction at coordinates $(x,y,t)$, the 3D $\gamma$-Index combines spatial displacement $(\Delta x,\Delta y)$, temporal offset $(\Delta t)$, and intensity deviation $(\Delta I)$ into a single normalized distance:
    \begin{equation}
    \gamma(x,y,t) =\min_{\substack{x',y' \in \Omega_{\texttt{DTA}} \\ t' \in \Omega_{\texttt{TTA}}}}\sqrt{\frac{(x-x')^2+(y-y')^2}{\texttt{DTA}^2} +
    \frac{(t-t')^2}{\texttt{TTA}^2} +\frac{\left(I_{\text{pred}}(x,y,t)-I_{\text{obs}}(x',y',t')\right)^2}{\texttt{IDT}^2}}\, .
    \end{equation}
The spatial and temporal search neighborhoods are defined as $\Omega_{\texttt{DTA}} = \left\{(x',y'):\ |x-x'|\le \frac{3}{2}\texttt{DTA},\ |y-y'|\le \frac{3}{2}\texttt{DTA}\right\}$,and $\Omega_{\texttt{TTA}} = \left\{t':\ |t-t'|\le \frac{3}{2}\texttt{TTA}\right\}$, which extend beyond the tolerance boundary to ensure that a feasible match is identified within operational margins. $I_{\text{pred}}$ and $I_{\text{obs}}$ denote predicted and observed irradiance. The intensity deviation tolerance \texttt{IDT} specifies the maximum acceptable irradiance deviation (for example 50~W/m$^2$), calibrated from sensor uncertainty or grid operational limits. Tolerance values should remain consistent with the native resolution of the forecast field. In practice, \texttt{DTA} is set to one grid spacing (one pixel). The temporal tolerance \texttt{TTA} can be limited to a half time step ($\Delta t/2$) when linear interpolation is used between consecutive images. The intensity tolerance \texttt{IDT} represents an acceptable relative deviation. A 10\% threshold is often suitable for radiative or energy quantities, but the value should be adapted to the variable and application. A prediction passes the criterion if $\gamma(x,y,t)\le 1$, where the threshold 1 corresponds to the boundary of the unit tolerance region in the normalized $\gamma$ space. The $\gamma$ passing rate \texttt{GPR} is computed as
    \begin{equation}
    \texttt{GPR} = \frac{1}{N}\sum_{i=1}^{N}\mathbbm{1}(\gamma_i \le 1)\times 100\%,
    \end{equation}
where $N=N_xN_yN_t$ and $\mathbbm{1}$ is the indicator function. The mean $\gamma$-Index is defined as $\bar{\gamma}=\frac{1}{N}\sum_{i=1}^{N}\gamma_i$, and the maximum $\gamma$-Index is $\gamma_{\max}=\max_{i\in\{1,\dots,N\}}\gamma_i$, which highlights localized important mismatches. The numerical minimization can be accelerated using the ellipse based evaluation illustrated in Figure~\ref{fig:gamma_ellipse}.}

\subsection{Tolerance Selection and Implementation}
\cyr{A key challenge in applying the $\gamma$-Index is the selection of appropriate \texttt{DTA}, \texttt{TTA}, and \texttt{IDT} values. Preliminary sensitivity tests indicate that a ±50 \% variation in \texttt{DTA}, \texttt{TTA}, or \texttt{IDT} modifies the $\gamma$-Passing Rate (\texttt{GPR}) by less than 10 \% and the mean $\gamma$ ($\bar{\gamma}$) by less than 0.3. These results hold for the present experimental conditions in Corsica and may vary with location, period, cloud regime, spatial and temporal resolutions or application specific constraints; overall, this suggests a limited dependence on the exact tolerance settings within the scope of this study. To ensure comparability across studies, this work adopts consistent tolerance parameters (\texttt{DTA}=1 pixel ($\approx 2$ km), \texttt{TTA}=30 min, \texttt{IDT}=50 W/m²), representative of realistic operational margins for high resolution forecasting. These values correspond to tolerances typically relevant at regional scales, where short term spatial and temporal variability is most significant. For finer or broader spatial domains, the same framework applies, but tolerance values should be adapted accordingly. An example of $\gamma$ calculation with synthetic \texttt{SSI} is given in Appendix~\ref{map} and the pseudo code for its computation is provided in Appendix~\ref{algo}. For completeness, Appendix~\ref{var} also details the different variants of the \texttt{RMSE} commonly used for spatio–temporal validation. These formulations (pixel wise, spatial, and aggregated) are not equivalent and may yield non unique or weakly interpretable results, making cross study comparisons unreliable unless the exact definition is explicitly stated.}
%
%
\section{Results}\label{Results}
\cyr{The section begin with a controlled numerical illustration and continue with the application to Corsica satellite fields. The analysis compares classical pixel based scores with the tolerance based $\gamma$ acceptance space and documents how each metric reacts to small spatio–temporal perturbations that preserve physical structure.}

\subsection{Synthetic Illustration}
\begin{figure}
        \centering
        \includegraphics[width=0.85\textwidth]{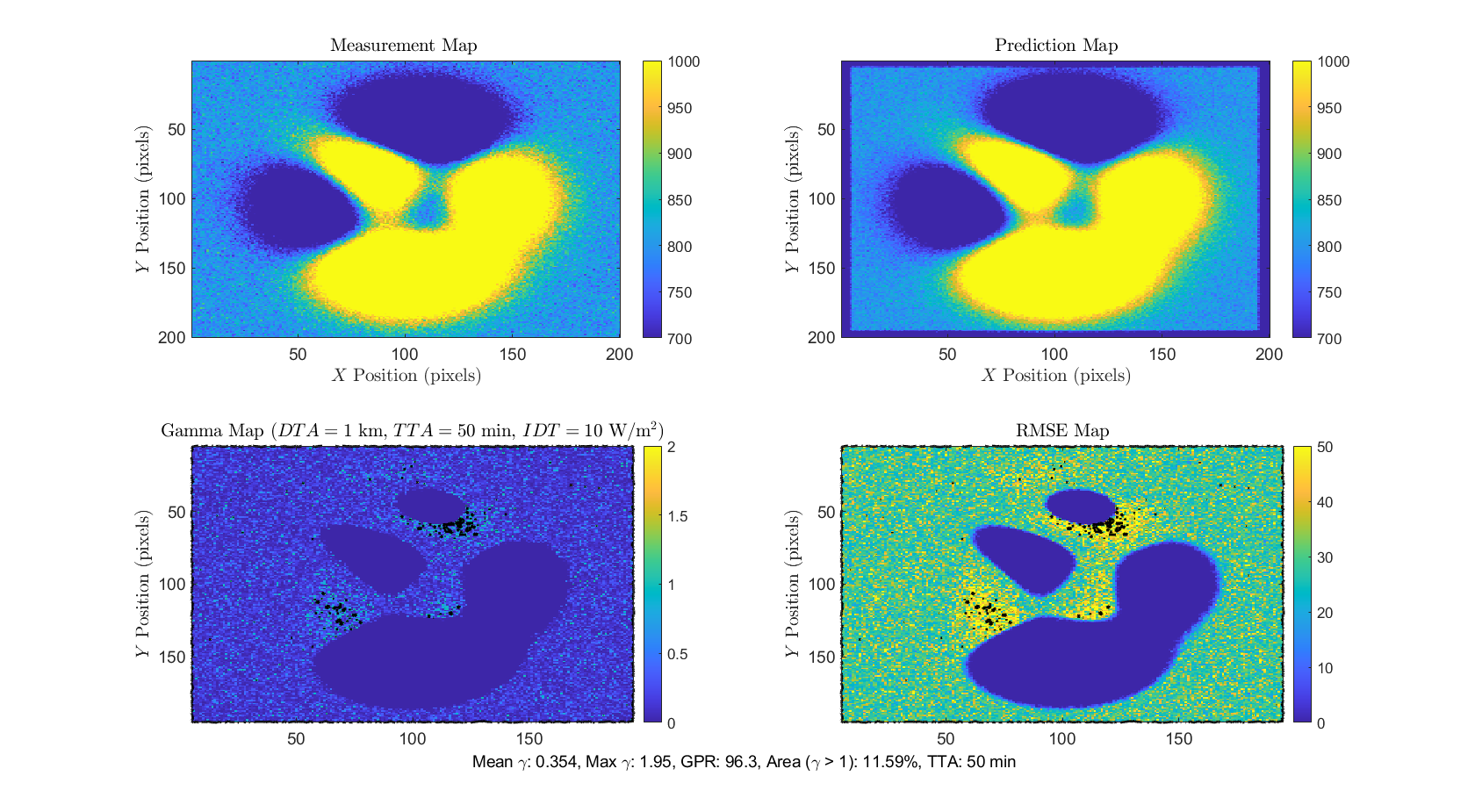}
        \caption{Comparison of mean pixel wise formulation of \texttt{RMSE} and the $\gamma$-Index for synthetic \texttt{SSI} validation. Black contours highlight areas where $\gamma>1$, indicating significant forecast errors.}
        \label{fig:gamma_map}
    \end{figure}

\cyr{Figure~\ref{fig:gamma_map} documents the behaviour of classical pixel wise \texttt{RMSE} and the local tolerance based $\gamma$ score for a controlled synthetic perturbation in $\mathbb{R}^3$ \citep{Low1998}. The experiment considers a reference field $I_{\text{obs}}(x,y,t)$ and a perturbed field $I_{\text{pred}}(x,y,t)=I_{\text{obs}}(x-\Delta x,y-\Delta y,t-\Delta t)+\epsilon(x,y,t),$ where $\Delta x \approx 1\,\texttt{pixel}$, corresponding to 2\,\texttt{km} in the studied grid, and $\Delta y=0$, $\Delta t=10\,\texttt{min}$, and $\epsilon$ is Gaussian noise. The resulting \texttt{RMSE\textsubscript{map}} yields uniformly high errors across the domain, suggesting poor skill despite preserved structure. In contrast, the $\gamma$ map shows that more than 80 percent of spatial pixels remain within the prescribed tolerances (\texttt{DTA}=2\,\texttt{km}, \texttt{TTA}=30\,\texttt{min}, \texttt{IDT}=50\,\texttt{W/m}$^2$); black contours indicate regions where $\gamma>1$, \textit{i.e.}, important mismatches.}

\subsection{HelioClim-3 Experiment}
\begin{figure}
        \centering
        \includegraphics[width=\textwidth]{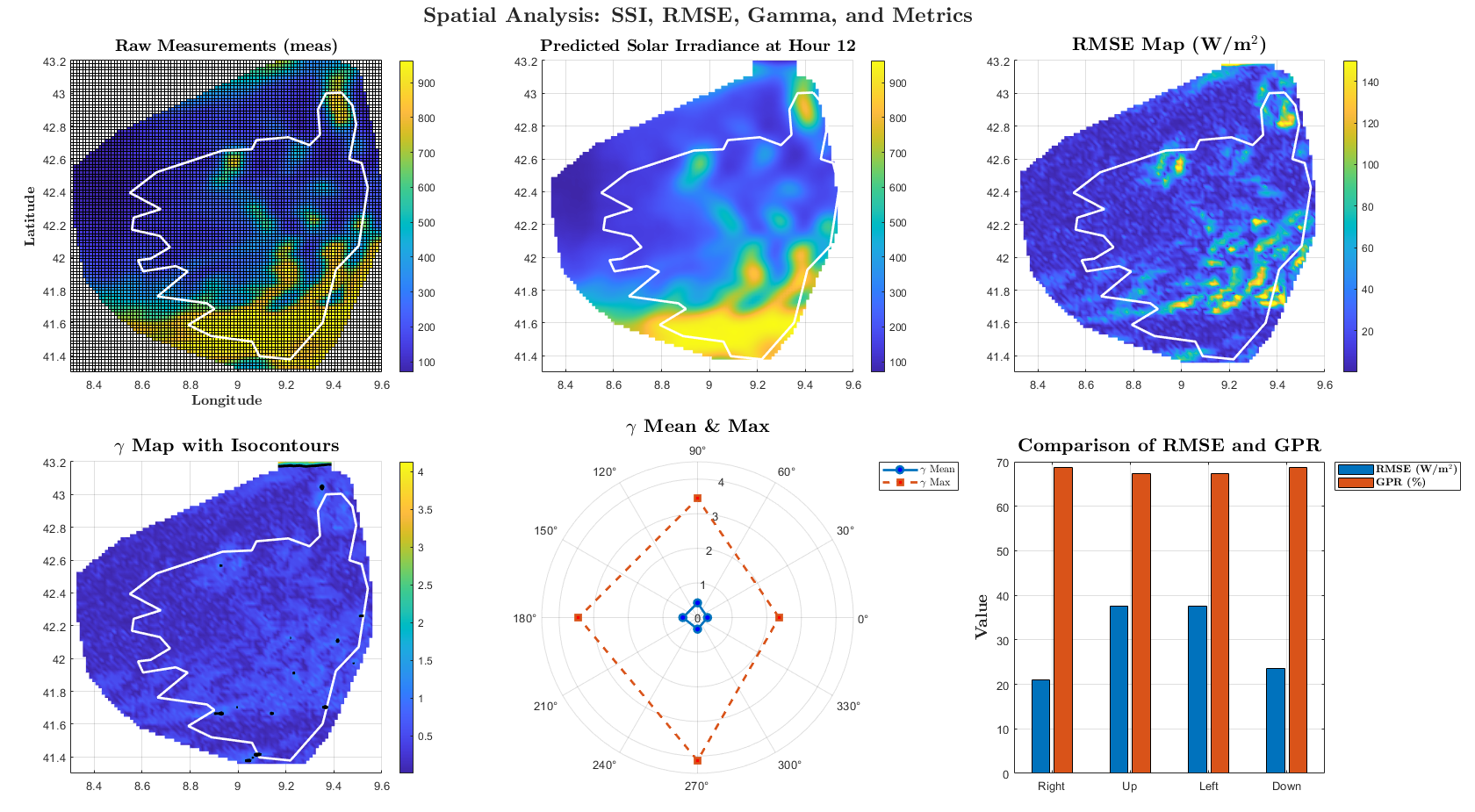}
        \caption{Impact of spatio-temporal displacements and Gaussian noise on solar irradiance metrics. The figure presents (a) raw \texttt{SSI} measurements (July 4, 2002), (b) predicted \texttt{SSI} after spatial and temporal shifts, (c) \texttt{RMSE} map, (d) $\gamma$-Index map highlighting tolerance thresholds, (e) radar plot comparing $\gamma$-mean and $\gamma$-max across shifts, and (f) bar chart comparing \texttt{RMSE} and \texttt{GPR} validation across perturbations.}
        \label{fig:analysis}
    \end{figure}

\cyr{The study uses satellite Surface Solar Irradiance fields from \texttt{HelioClim-3} \citep{rs70709269} and from \texttt{CAMS Radiation Service} \citep{asr-17-143-2020}. Forecast maps are evaluated relative to the observed reference map $I_{\text{obs}}(x,y,t)$ through a map to map workflow consistent with the native spatial grid~$\Omega$ and time sequence. The acceptance assessment applies the 3D $\gamma$ Index with prescribed tolerances \texttt{DTA}=2\,km, \texttt{TTA}=30\,min, and \texttt{IDT}=50\,W/m$^2$ \citep{Low1998}. Figure~\ref{fig:analysis} compares the local pixel aggregation of $\gamma$ with the classical pixel wise formulation of \texttt{RMSE\textsubscript{map}}, and summarizes the actual experiment under controlled perturbations. Pixel wise \texttt{RMSE\textsubscript{map}} for the studied situations varies between 20.95 and 37.65\,W/m$^2$, which reflects a strong sensitivity to spatial and temporal misalignment of coherent gradients \citep{YANG202020,Casati2004,Roberts2008,Gilleland2009}. In contrast, the mean $\gamma$ value remains stable between 0.29 and 0.35, and the \texttt{GPR} between 67.35 and 68.64~\%, which expresses robustness to small displacements embedded in the tolerance philosophy \citep{Low1998}. The maximum $\gamma$ identifies localized regions where $\gamma>1$, highlighting important mismatches not visible in domain averages; this localization property is consistent with neighbourhood and multiscale verification research \citep{Roberts2008,Casati2004,Gilleland2009}. Systematic deviations are described through \texttt{MBE} oscillating from $-7.2$ to $3.6$\,W/m$^2$; these bias values remain essential for solar engineers to assess monotonic field shifts even when $\bar{\gamma}$ and \texttt{GPR} are acceptable \citep{YANG202020}. For clear and fully overcast regimes, $\gamma$ and \texttt{RMSE} provide similar qualitative conclusions because irradiance fields are smooth and weakly structured, a behaviour already reported in spatial acceptance studies \citep{Roberts2008,Casati2004,Gilleland2009}. The largest differences arise under intermediate and broken cloud situations, where sharp spatial gradients amplify point wise penalties: \texttt{RMSE\textsubscript{map}} over penalizes small coherent displacements, while $\gamma$ maintains a joint acceptance rule in $\mathbb{R}^3$ capable of distinguishing operationally negligible offsets from genuine field discrepancies, as originally motivated in medical physics verification \citep{Low1998}.}
%
%
\section{Discussion}\label{sec:discussion}
Results suggest that the $\gamma$-Index provides a more meaningful metric for evaluating \texttt{SSI} forecasts compared to \texttt{RMSE}, offering a more physics aware validation approach that aligns better with operational constraints. The $\gamma$-Index follows the same rationale as spatial verification approaches \citep{Gilleland2009}, but formulates tolerances directly in physical units of space, time, and intensity, which makes it easier to interpret in operational contexts.
Beyond its technical interpretation, the $\gamma$-Index helps grid operators avoid both over and undercorrections related to forecast misalignment. For instance, if a cloud forecast is spatially offset by 3\,km and temporally by 15\,minutes, \texttt{RMSE} would suggest failure, while $\gamma$ would validate the forecast within defined tolerances. This difference is non trivial: overestimation of forecast error may lead operators to trigger backup systems (\textit{e.g.}, gas turbines or batteries) unnecessarily. A $\gamma$-based validation enables more trust in forecasts exhibiting small but acceptable deviations, which may result in cost savings and improved system resilience. From a methodological perspective, the mathematical structure of the $\gamma$-Index also opens perspectives beyond forecast verification. 
Although the present study focuses on validation, the continuous and bounded nature of $\gamma$ suggests its possible use as a constraint in optimization frameworks. 
Because $\gamma(x,y,t)$ is continuous and differentiable almost everywhere (except at transition points where the minimal correspondence changes) it can be used in gradient based optimization. The non differentiability arises only from the $\max(0,\,\gamma-1)$ term, analogous to the \texttt{ReLU} function commonly used in deep learning losses:
\begin{equation}
\mathcal{L}_{\text{total}} = \mathcal{L}_{\text{data}} + \lambda \, \mathbb{E}\!\left[\max(0,\, \gamma(x,y,t) - 1)\right],
\end{equation}
where $\mathcal{L}_{\text{data}}$ denotes the standard reconstruction or prediction loss (\textit{e.g.}, mean squared or absolute error) and $\lambda$ controls the strength of the spatial–temporal tolerance constraint. 
Such a formulation would penalize predictions exceeding acceptable deviations (\textit{i.e.}, $\gamma > 1$), the offset ``$-1$'' corresponding exactly to the physical tolerance boundary of the $\gamma$-Index, while maintaining differentiability for gradient based optimization. This prospect opens a potential path toward physics (and tolerance) aware learning schemes, but remains beyond the scope of the present validation work.
%
%
\section{Limitations}\label{limitations}
The present formulation intentionally favours simplicity and numerical robustness. Fixed tolerances were adopted (\texttt{DTA}=1\,pixel, \texttt{TTA}=\texttt{$\Delta$t}/2, \texttt{IDT}=10\%), which provide a stable reference for typical \texttt{SSI} or \texttt{NWP} resolutions (1–5\,km, hourly). However, several limitations arise from these assumptions. First, the temporal component was not fully exploited in this first implementation to avoid the computational cost and potential artefacts of temporal interpolation; the validation thus primarily focuses on the spatial dimension. A finer \texttt{TTA} or a full temporal search would be required to capture transient phenomena or to discriminate persistence models, which can otherwise be favoured if \texttt{TTA} equals the full sampling interval \texttt{$\Delta$t}. Second, setting \texttt{DTA}=1\,pixel is appropriate for moderate resolutions, but for coarser grids (\textit{e.g.}, $\Delta x > 5$–10\,km) this tolerance would exceed the physical coherence scale of cloud fields.  In such cases, aggregation or scale adaptive weighting of neighbouring pixels would be required to maintain spatial relevance. Finally, the 10\% intensity tolerance (\texttt{IDT}) is a convenient baseline, but the operational rules of the variable (irradiance, wind, precipitation, or temperature) should dictate the acceptable deviation threshold.
%
%
\section{Conclusions}\label{sec:Conclusions}  
The $\gamma$-Index provides a unified and physically grounded framework for the spatio–temporal validation of forecast fields. By combining spatial (\texttt{DTA}), temporal (\texttt{TTA}), and intensity (\texttt{IDT}) tolerances, it quantifies forecast quality in a way that explicitly accounts for displacement and amplitude uncertainties. Unlike pixel wise error metrics such as \texttt{RMSE} or \texttt{MAE}, which over penalize small spatial or temporal shifts and under penalize broad structural or bias related discrepancies, the $\gamma$-Index preserves the operational meaning of forecast errors and better reflects actionable accuracy. Applied here to \texttt{SSI} maps, the method demonstrates how tolerance based validation can provide a more stable and interpretable assessment of predictive skill. The derived $\gamma$-Passing Rate (\texttt{GPR}) and mean $\gamma$ offer concise indicators for model comparison, post processing evaluation, and operational confidence building. Thresholds such as $\texttt{GPR}>80\%$ and $\bar{\gamma}<0.7$ provide practical guidance for assessing forecast reliability under realistic conditions. Beyond solar applications, the same formulation can be extended to any gridded forecast variable (temperature, wind, precipitation, or pollutant concentration) provided appropriate tolerances are defined. This generality makes the $\gamma$-Index a candidate tool for harmonizing spatial–temporal verification across domains. 

While established methods such as the Fractions Skill Score \citep{Roberts2008} and object based approaches like MODE \citep{davis2006} or SAL \citep{aaaa} provide complementary scale dependent and feature level diagnostics, the $\gamma$-Index offers explicit specification of independent physical tolerances in three dimensions, suggesting potential for hybrid verification frameworks that combine scale aware skill assessment with tolerance based operational compliance criteria. 

Future work should therefore focus on adaptive and data driven selection of tolerance parameters (\texttt{DTA}, \texttt{TTA}, and \texttt{IDT}), formal bias and sensitivity analysis, and the integration of temporal interpolation strategies. Further studies could also explore its use within learning frameworks as a tolerance based regularization term, and assess its impact on operational decisions, such as grid balancing or dispatch optimization. Although demonstrated here on Surface Solar Irradiance, the proposed tolerance based validation framework is directly transferable to other geophysical forecasts (\textit{e.g.}, precipitation, wind, or temperature), as long as relevant \texttt{DTA}, \texttt{TTA}, and \texttt{IDT} values are defined. Overall, the $\gamma$-Index complements existing verification metrics by bridging physical relevance and statistical rigor, providing a pragmatic path toward more consistent and interpretable forecast evaluation.
%
%

\section*{Appendix}
\appendix
\section{Beyond \texttt{RMSE} or \texttt{MAE}: A Spatiotemporal Perspective on Forecast Validation}
\label{map}
To clarify the scope of our comparison and the formulation of reference metrics, the main variants of the \texttt{RMSE} used in map based forecast evaluation are detailed before illustrating their behaviour relative to the $\gamma$-Index. The well known ``double penalty'' effect mainly occurs when the mean pixel wise \texttt{RMSE} over a forecast map is considered. In such a case, a simple spatial or temporal displacement of a coherent structure (\textit{e.g.}, a cloud field) induces two simultaneous errors (a miss and a false alarm) thus inflating the global score. This issue is important for high resolution nowcasts and local grid applications, where pointwise accuracy is less relevant than structural consistency. For coarse or aggregated evaluations, such displacements have negligible impact, and classical metrics remain appropriate. Alternative formulations exist, including \texttt{MAE} to reduce sensitivity to outliers, \texttt{RMSE} of spatially averaged fields, or ensemble based probabilistic scores. However, these approaches either lose spatial interpretability or omit explicit physical tolerances. The $\gamma$-Index preserves map-level consistency by integrating distance, time, and intensity tolerances within a single normalized criterion.

\section{Variants of Classical Error Metrics in 2D–3D Forecast Workflows}
\label{var}
For spatio–temporal forecasts such as \texttt{SSI} maps (or other map), several non equivalent formulations of the Root Mean Square Error (\texttt{RMSE}) are commonly used, leading to ambiguous interpretations. Pixel wise temporal \texttt{RMSE}. Each pixel $(i,j)$ is considered as a temporal series of irradiance values, the definition using the $L^1$-norm (\textit{i.e.}, \texttt{MAE}) is analogous:
\begin{equation}
\texttt{RMSE}_{ij} =
\sqrt{\frac{1}{N_t}\sum_{t=1}^{N_t}(\hat{I}_{ij,t}-I_{ij,t})^2},
\end{equation}
and the map-level score is the spatial mean:
\begin{equation}
\texttt{RMSE}_{\mathrm{map}} =
\frac{1}{N_x N_y}\sum_{i,j}\texttt{RMSE}_{ij}.
\end{equation}
This form, used in the present work, suffers from the double penalty effect, where a small spatial or temporal shift of a cloud induces two simultaneous errors. The Instantaneous spatial \texttt{RMSE} is different, at each time step $t$:
\begin{equation}
\texttt{RMSE}_t =
\sqrt{\frac{1}{N_x N_y}\sum_{i,j}(\hat{I}_{ij,t}-I_{ij,t})^2},
\end{equation}
and the temporal mean gives $\texttt{RMSE}_{\mathrm{time}} = \frac{1}{N_t}\sum_t \texttt{RMSE}_t$. This variant focuses on spatial fidelity but ignores temporal alignment. The last variant is the \texttt{RMSE} of the spatially averaged field.
\begin{equation}
\texttt{RMSE}_{\mathrm{avg}} =
\sqrt{\frac{1}{N_t}\sum_t(\overline{\hat{I}}_t-\overline{I}_t)^2}, \qquad
\overline{I}_t = \frac{1}{N_x N_y}\sum_{i,j}I_{ij,t}.
\end{equation}
This aggregated measure captures global bias but loses spatial structure. Each of these forms can be normalized as $\texttt{nRMSE}=\texttt{RMSE}/\overline{I}$, where $\overline{I}$ may denote a spatial, temporal, or global mean irradiance, introducing further inconsistency across datasets and studies.  Because these formulations emphasize distinct aspects (temporal coherence, spatial accuracy, or global bias), \texttt{RMSE} cannot serve as a single, physically interpretable metric for spatio–temporal validation. The $\gamma$-Index overcomes this limitation by jointly integrating spatial (\texttt{DTA}), temporal (\texttt{TTA}) and intensity (\texttt{IDT}) tolerances into a single normalized criterion.

\section{Pseudo Code for $\gamma$-Index Computation}  
\label{algo}
This appendix presents an algorithm for computing the $\gamma$-Index alongside traditional error metrics such as Root Mean Square Error (\texttt{RMSE}) and $\gamma$ Passing Rate (\texttt{GPR}). The approach incorporates spatial tolerance (\texttt{DTA}, km), temporal tolerance (\texttt{TTA}, min), and intensity difference tolerance (\texttt{IDT}, W/m$^2$) to enable precise comparison between predicted and observed data.  
The function inputs include a predicted irradiance map (\texttt{pred}, $n_y \times n_x$, W/m$^2$), a time sequenced set of measured maps (\texttt{meas}, $n_y \times n_x \times n_t$, W/m$^2$), initial and final pixel resolutions (\texttt{res\_ini}, \texttt{res\_end}, km), and initial and target time intervals between images (\texttt{timestep\_ini}, \texttt{timestep\_end}, hours). Algorithm \ref{alg:gamma} consist of outputs related to the $\gamma$-Index Map (sized by $n_y \times n_x$), mean and max $\gamma$ values coupled with The $\gamma$ Passing Rate (\texttt{GPR}), representing the percentage of pixels where $\gamma \leq 1$. As example, the \texttt{RMSE} map is also presented.  
\begin{algorithm}
\caption{$\gamma$-Index and \texttt{RMSE} Computation}
\label{alg:gamma}
\begin{algorithmic}[1]
\Require $\texttt{pred}$ $(n_y, n_x)$, $\texttt{meas}$ $(n_y, n_x, n_t)$, $\texttt{DTA}, \texttt{TTA}, \texttt{IDT}$, $\texttt{res\_ini}, \texttt{res\_end}$, $\texttt{timestep\_ini}, \texttt{timestep\_end}$  
\Ensure $\texttt{gammaMap}, \texttt{gammaMean}, \texttt{gammaMax}, \texttt{RMSE}, \texttt{GPR}, \texttt{rmseMap}$  
\State \textbf{Adjust Spatial and Temporal Resolution}  
\If {$res\_ini \neq res\_end$} \State Resize $\texttt{pred}$ and $\texttt{meas}$ \EndIf  
\If {$timestep\_ini \neq timestep\_end$} \State Interpolate $\texttt{meas}$ in time \EndIf  
\State \textbf{Initialize Outputs}  
\State $\texttt{gammaMap} \gets \infty$, $\texttt{rmseMap} \gets 0$  
\State \textbf{Compute $\gamma$ and \texttt{RMSE}}  
\For {$i, j$ in grid}  
    \State $d_{\max} \gets 3\texttt{DTA} / (2 \times res\_end)$, $t_{\max} \gets 3\texttt{TTA} / (2 \times timestep\_end \times 60)$  
    \State Extract local region from $\texttt{meas}$ within $d_{\max}$, $t_{\max}$  
    \State Compute $\gamma = \min \sqrt{(j - X)^2 / \texttt{DTA}^2 + (i - Y)^2 / \texttt{DTA}^2 + (t - T)^2 / \texttt{TTA}^2 + (pred(i,j,t) - meas(i,j,t))^2 / \texttt{IDT}^2}$  
    \State $\texttt{gammaMap}(i,j) \gets \gamma$, $\texttt{rmseMap}(i,j) \gets \sqrt{\frac{1}{n_t} \sum_{t} (pred(i,j) - meas(i,j,t))^2}$  
\EndFor  
\State \textbf{Compute Global Metrics}  
\State $\texttt{gammaMean} \gets \text{mean}(\texttt{gammaMap})$, $\texttt{gammaMax} \gets \text{max}(\texttt{gammaMap})$, $\texttt{RMSE} \gets \text{mean}(\texttt{rmseMap})$, $\texttt{GPR} \gets \frac{\text{count}(\texttt{gammaMap} \leq 1)}{n_y n_x} \times 100$
\end{algorithmic}
\end{algorithm}
%
%
\bibliographystyle{cas-model2-names}
\bibliography{gamma}
\end{document}